\documentclass{article}
\usepackage{amsfonts}
\usepackage{float}
\usepackage{amsmath}
\usepackage{bm}
\usepackage{cite}
\usepackage{graphicx}

\hfuzz5pt \allowdisplaybreaks \numberwithin{equation}{section}
 \DeclareMathOperator{\sgn}{\rm sgn}
 \DeclareMathOperator{\diag}{\rm diag}
 
 \DeclareMathOperator*{\res}{\rm res}

 \DeclareMathOperator{\Rs}{\mathbb{R}}
 \DeclareMathOperator{\Cs}{\mathbb{C}}

 \DeclareMathOperator{\Ao}{\mathcal{A}}
 \DeclareMathOperator{\Lo}{\mathcal{L}}

 \DeclareMathOperator{\Fo}{\mathcal{F}}
 \DeclareMathOperator{\Do}{\mathcal{D}}
 
 \DeclareMathOperator{\Ncal}{\mathcal{N}}
 \DeclareMathOperator{\So}{\mathcal{S}}

 \DeclareMathOperator{\q}{\mathbf{q}}
 \DeclareMathOperator{\bk}{\mathbf{k}}

\begin{document}

\title{Building extended resolvent of heat operator via twisting transformations}
\author{M.~Boiti${}^{*}$, F.~Pempinelli${}^{*}$, A.~K.~Pogrebkov${}^\dag$,
and B.~Prinari${}^{*}$ \\
{\ }\\
${}^{*}$Dipartimento di Fisica, Universit\`a del Salento and\\
Sezione INFN, Lecce, Italy\\
${}^\dag$Steklov Mathematical Institute, Moscow, Russia}
\date{\ }
\maketitle

\begin{abstract}
Twisting transformations for the heat operator are introduced. They are used, at the same time, to superimpose \`{a} la Darboux $N$ solitons to a generic
smooth, decaying at infinity, potential and to generate the corresponding Jost solutions. These twisting operators are also used to study the existence of the
related extended resolvent. Existence and uniqueness of the extended resolvent in the case of $N$ solitons with $N$ ``ingoing'' rays and one
``outgoing'' ray is studied in details.
\end{abstract}

\section{Introduction}

The Kadomtsev--Petviashvili equation in its version called KPII
\begin{equation}
 (u_{t}-6uu_{x_{1}}+u_{x_{1}x_{1}x_{1}})_{x_{1}}=-3u_{x_{2}x_{2}},\label{KPII}
\end{equation}
is a (2+1)-dimensional generalization of the celebrated Korteweg--de~Vries (KdV) equation. As a consequence, the KPII equation admits solutions that
behave at space infinity like the solutions of the KdV equation. For instance, if $u_{1}(t,x_{1})$ obeys KdV, then
$u(t,x_{1},x_{2})=u_{1}(t,x_{1}+\mu x_{2}-3\mu ^{2}t)$ solves KPII for an arbitrary constant $\mu \in \mathbb{R}$. Thus, it is important to consider
solutions of~(\ref{KPII}) that decay at space infinity in all directions with exception of a finite number of 1-dimensional rays with behavior of the type of $u_{1}$. Even though KPII has been known to be integrable for more than three decades~\cite{D1974,ZS1974}, its general theory (involving such non-decaying
solutions) is far from being complete. Thus, the Cauchy problem for KPII with rapidly decaying initial data was solved in~\cite{AYF1983,Grinev1988}
by using the Inverse Scattering Transform (IST) method, based on the spectral analysis of the heat operator
\begin{equation}
 \Lo(x,\partial_{x}^{})=-\partial_{x_{2}}^{}+\partial_{x_{1}}^{2}-u(x),\qquad x=(x_{1}^{},x_{2}^{}),  \label{heatop}
\end{equation}
that gives the associated linear problem for the KPII equation. The standard approach to the spectral theory of the operator~(\ref{heatop}) is
based on integral equations for the Jost solution $\Phi(x,\bk)$, where $\bk\in\Cs$ denotes the spectral parameter, or for $\Psi(x,\bk)$, Jost solution of the dual operator ${\Lo}^{\text{d}}$. However, it is known that in the case of potentials with one-dimensional asymptotic behavior these integral
equations are ill-defined.

In order to overcome these difficulties, a resolvent approach was developed
in~\cite{BPPPo1992a,BPPPo1992c,BPP1994c,BPP1997,BPPPr1998,Pr2000,BPPPr2001a,BPPPr2002,BPPPr2005a,BPPPr2005b}. In this approach a space of operators
$A(q)$ with kernels $A(x,x';q)$ belonging to the space of tempered distributions of variables $x,x',q\in\Rs^{2}$ was introduced. Ordinary
differential operators $\Lo(x,\partial_{x})$ are imbedded in this space as operators with kernels
\begin{equation}
 L(x,x';q)\equiv \Lo(x,\partial_{x}+q)\delta (x-x').\label{2}
\end{equation}
Due to the additional dependence on the parameter $q$ these operators are called extended operators. In this space a generalization of the
resolvent of a differential operator, called extended resolvent, was introduced, enabling consideration of the spectral theory of operators with
nontrivial asymptotic behavior at space infinity.

In~\cite{BPPPr2001a,BPPPr2002,BPPPr2005a,BPPPr2005b} we considered the nonstationary Schr\"{o}dinger and heat operators with potentials with only one direction of nondecaying behavior. The first step in solving the problem was the embedding of a pure one-dimensional spectral theory in the
two-dimensional one, building the two-dimensional extended resolvent for an operator with potential $u(x)\equiv u_{1}(x_{1})$. The second step was the
dressing of this resolvent by an arbitrary bidimensional perturbation of the potential $u_{1}$. Finally, all mathematical entities appearing in the
IST theory were obtained from this dressed resolvent by a reduction procedure, allowing the formulation of the direct and inverse problems.  Let us also mention that the standard spectral theory for the heat operator in the case of potentials nondecaying in one space direction was developed in~\cite{Villaroel}, but under some special and inexplicit conditions on the potential.

Here we consider the substantially more complicated problem: the case of a potential $u$ non decaying along multiple non parallel rays. Thus now there is no analogy with the one-dimensional case and the whole theory has to be constructed directly, without embedding one-dimensional entities in two dimensions. Therefore, we are obliged to consider directly true bidimensional potentials as it was already done in~\cite{BPP2006a,BPP2006b} for the nonstationary Schr\"{o}dinger operator. In fact, we apply the same procedure that was successfully applied in that case. Precisely, in order to get the potential corresponding to $N$ solitons ``superimposed'' to a generic smooth decaying potential and the related Jost solutions we construct these entities directly by means of twisting transformations instead of using recursively the (binary) Darboux transformation. In this way we recover potentials and Jost solutions that were obtained in~\cite{BPPPr2001a} by using a recursive procedure and, also, some alternative representations of the pure solitonic potentials obtained in \cite{MZBIM1977,BK,BC,BC2,ChK1} by using the $\tau $ functions. It is worth noting that the heat operator is not self-dual and thus singular behaviors of the left and right twisting operators are not obliged to be correlated. This explains why the structure of the $N$ soliton solution of KPII is much richer than for KPI. In particular, the $N$ soliton solution can have a different number of ``ingoing'' and ``outgoing''  (in the sense of $x_2\to-\infty$, or $+\infty$) rays.

The article is organized as follows. In Sec.~\ref{back} we briefly review some aspects and basic ideas of the extended resolvent approach,
referring to the example of an operator~(\ref{heatop}) with a smooth, rapidly decaying potential $u(x)$. For further details, we refer the
interested readers to~\cite{BPPPo1992a,BPPPo1992c,BPP1994c,BPP1997,BPPPr1998}. In Sec.~\ref{twist} we introduce the twisting operators, i.e.,
operators that ``twist'' the operator $L$, extension in the sense of~(\ref{2}) of the operator $\Lo$ in~(\ref{heatop}), to an operator $L'$ of the same
kind with a potential $u'$ describing $N$ solitons ``superimposed'' \`a la Darboux to the background potential $u$. In Sec.~\ref{resolvent}, by using these
twisting operators, under the assumption of the existence of the resolvent $M'(q)$ we derive an explicit expression for its kernel $M'(x,x';q)$. So the problem of the existence of the resolvent $M'(q)$ is reduced to the problem of finding the region in the $q$-plane, where this kernel is a tempered distribution. This problem resulted to be more arduous than in the case of the nonstationary Schr\"{o}dinger operator and, therefore, in Sec.~\ref{soliton}, we first consider the pure $N$-soliton potentials and, afterwards, in Sec.~\ref{1ray}, the special simple subclass of $N$-soliton potentials with $N$ ``ingoing'' rays and only one
``outgoing'' ray. We prove that for $N>1$ the resolvent $M'(q)$ exists only in the region of the $q$-plane exterior to a polygon with $N+1$ sides. More specifically, we prove that there exists a value $\bk_0$ of the spectral parameter $\bk$, for which the dual Jost solution $\Psi'(x,\bk)$ is such that $\Psi'(x,\bk_0)\exp(q_1x_1+q_2x_2)$  is exponentially decaying on the $x$-plane for any value of $q$ belonging to this polygon and, consequently, we deduce that inside the polygon the twisted operator $L'(q)$ has a left annihilator and cannot have right inverse. Nonetheless, the Green's function of $\Lo'$ exists and can be uniquely derived via a reduction from $M'(q)$. In a forthcoming publication we plan to consider the generic case of an arbitrary number of incoming and outgoing rays and to elucidate the role of these annihilators in the spectral theory of such potentials.

\section{Background theory}\label{back}

Let us introduce the space of \textbf{extended} operators $A(q)$, i.e., operators with kernel $A(x,x';q)$ belonging to the space $\So'$ of tempered distributions of the six real variables $x=(x_{1},x_{2})$, $x'=(x_{1}',x_{2}')$, and
$q=(q_{1},q_{2})$. For two extended operators $A(q)$ and
$B(q)$ with kernels $A(x,x';q)$ and $B(x,x';q)$ we introduce the composition law
\begin{equation}
 (AB)(x,x';q)=\int dx''\,A(x,x'';q)\,B(x'',x';q),  \label{3}
\end{equation}
provided the integral exists in terms of distributions. An operator $A$ can have an inverse, $A^{-1}$, in the sense of this composition: $AA^{-1}=I$ and $A^{-1}A=I$, where $I$ is the unity operator, i.e., the operator with kernel $I(x,x';q)=\delta(x_{1}-x'_1)\delta(x_{2}-x'_2)$. A specific subclass of extended operators is given by the extensions $L(q)$ of differential operators as defined in~(\ref{2}), where $\Lo(x,\partial_{x})$ denotes a differential operator whose coefficients are smooth functions of $x$. Let us associate to any operator $A(q)$ with kernel $A(x,x';q)$ its ``hat''-kernel
\begin{equation}
 \widehat{A}(x,x';q)=e_{}^{q(x-x')}A(x,x';q), \label{7}
\end{equation}
where $qx=q_{1}x_{1}+q_{2}x_{2}$. For a differential operator $L(q)$ this procedure is the inverse of the extension introduced
in~(\ref{2}), i.e., $\widehat{L}(x,x';q)=L(x,x')$. Notice, however, that, since the kernels of the operators $A$ are only subjected to the
requirement of belonging to the space of tempered distributions, in general, the hat-kernel $\widehat{A}$ still depends on $q$ and is not necessarily
bounded. The conjugate $A^{\ast}(q)$ of an operator $A(q)$ is defined by
\begin{equation}
 A^{\ast}(x,x';q)=\overline{A(x,x';q)}. \label{adg}
\end{equation}

It is convenient to introduce the representation of the operator $A(q)$ in the $p$-space which is defined by the Fourier transform
\begin{equation}
 A(p;\q)=\dfrac{1}{(2\pi)^{2}}\int dx\int dx'\,e^{i(p+\q_{\Re})x-i\q_{\Re}x'}A(x,x';\q_{\Im}),  \label{px}
\end{equation}
where $p=(p_{1},p_{2})\in\Rs^{2}$ and we introduced a two dimensional \textbf{complex} vector
\begin{equation}
 \q=\q_{\Re}+i\q_{\Im},\qquad q\equiv \q_{\Im},\quad \q_{\Re},\q_{\Im}\in \Rs^{2}.  \label{qbold}
\end{equation}
The composition (\ref{3}) in the $p$-space becomes a sort of shifted convolution
\begin{equation}
 (AB)(p;\q)=\int dp'A(p-p';\q+p')B(p';\q). \label{comp}
\end{equation}

Then, with these notations, the extension of the heat operator~(\ref{heatop}) reads
\begin{equation}
 L=L_{0}-u,\qquad u(x,x';q)=u(x)\delta (x-x'),  \label{L}
\end{equation}
where $L_0$ is the extension in the sense of~(\ref{2}) of the differential part $\Lo_0$ of the heat operator~(\ref{heatop}) and has, in the $p$-space, thanks to~(\ref{px}), kernel given by
\begin{equation}\label{L0}
    L_0(p;\q)=(i\q_2^{}-\q_1^2)\delta(p).
\end{equation}
The main object of our approach is the extended resolvent (or resolvent, for short) $M(q)$ of the operator $L(q)$, which is the inverse of
the operator $L$ in the sense of composition~(\ref{3}) (or~(\ref{comp})), i.e.,
\begin{equation}
 LM=ML=I,  \label{M}
\end{equation}
and can be defined as the solution of the integral equations
\begin{equation}
 M=M_{0}+M_{0}uM,\qquad M=M_{0}+uMM_{0},  \label{inteqM}
\end{equation}
where $M_0$ is the resolvent of the zero potential (bare) operator $L_0$. Thanks to~(\ref{L0}) we have $M_0(p;\q)=\delta(p)/(i\q_2-\q^2_1)$ for the
kernel of this operator in the $p$-space. In the case of a rapidly decaying potential $u(x)$, the existence and uniqueness of the solution
of the equations in~(\ref{inteqM}) can be proved in analogy with~\cite{Grinev1988}. Notice that for a real potential $u(x)$ both $L$ and $M$ are
self-conjugate operators in the sense of definition~(\ref{adg}).

Thanks to~(\ref{comp}) and the explicit form of $M_0(p;\q)$ given above the integral equations~(\ref{inteqM}) written in the $p$-space show that the
kernel $M(p;\q)$ is singular for $\q_2=-i\q_1^2$ and for $\q_2+p_2=-i(\q_1+p_1)^2$. Therefore, it is natural to introduce the following truncated and
reduced values of the resolvent
\begin{equation}
 \nu (p;\q)=(ML_{0})(p;\q)\Bigr|_{\q_2=-i\q_1^2},\quad
 \omega(p;\q)=(L_{0}M)(p;\q)\Bigr|_{\q_2=-i(\q_1+p_1)^2-p_2}.  \label{H.3.103}
\end{equation}
One can show that the operator $L(q)$ and its resolvent $M(q)$ admit the following bilinear representations in terms of $\nu$ and $\omega$
\begin{equation}
 L=\nu L_{0}\omega ,\qquad M=\nu M_{0}\omega .  \label{LM}
\end{equation}
Correspondingly, we call the operators $\nu$ and $\omega$ dressing operators since they ``dress'' the bare operators $L_{0}$ and $M_{0}$. Notice that
$\nu(p;\q)$ and $\omega(p;\q)$ satisfy asymptotics $\lim_{\q_{1}\to\infty}\nu (p;\q)=\delta (p)$ and $\lim_{\q_{1}\to\infty}\omega(p;\q)=\delta (p)$ and
do not depend on $\q_{2}$, which, if necessary, we make clear by writing $\nu (p;\q_{1})$ and $\omega (p;\q_{1})$.

The dressing operators $\nu$ and $\omega$ satisfy the equations
\begin{equation}
 L\nu =\nu L_{0},\qquad \omega L=L_{0}\omega ,  \label{Lnu}
\end{equation}
and are mutually inverse
\begin{equation}
\omega\nu =I,\qquad \nu\omega =I.  \label{scalar}
\end{equation}

In order to define the Jost solutions by means of the dressing operators we introduce the following Fourier transforms:
\begin{equation}
 \chi (x,\q_1)=\int dp\,e^{-ipx}\nu(p;\q_1),\qquad \xi (x,\q_1)=\int dp\,e^{-ipx}\omega (p;\q_1-p_{1}).  \label{nuchi}
\end{equation}
Then the Jost and the dual Jost solutions can be defined, respectively, as
\begin{equation}
 \Phi (x,\bk)=e^{-i\bk x_1-\bk^2x_2}\chi (x,\bk),\qquad \Psi(x,\bk)=e^{i\bk x_1+\bk^2x_2}\xi (x,\bk),  \label{dualJost}
\end{equation}
where we denoted $\q_1$ as $\bk$ in order to meet the standard notations for the spectral parameter . Thanks to~(\ref{Lnu}), they
obey the heat equation and its dual
\begin{equation}
 (-\partial_{x_{2}}^{}+\partial_{x_{1}}^{2}-u(x))\Phi (x,\bk)=0,\quad (\partial_{x_{2}}^{}+\partial_{x_{1}}^{2}-u(x))\Psi (x,\bk)=0. \label{eq-Phi}
\end{equation}

\section{Darboux transformations via twisting operators $\zeta$ and $\eta$}\label{twist}

In order to build a two-dimensional potential describing $N$ solitons superimposed to a generic smooth background, we bypass the recursive procedure
used in \cite{BPPPr2001a} and, by using the operator formulation introduced in the previous section, we construct directly the final Darboux
transformation. The main tool in accomplishing this result is what we call the twisting operators.

Let us consider a transformation from the operator $L$ in~(\ref{L}) to a new operator of the same form
\begin{equation}
 L'=L_{0}-u',\qquad u'(x,x';q)=u'(x)\delta(x-x'),  \label{tildeL}
\end{equation}
given by means of an operator pair $\zeta$, $\eta$ according to the formulas
\begin{equation}
 L'\zeta =\zeta L,\qquad \eta L'=L\eta ,  \label{Lzeta}
\end{equation}
``twisting'' $L$ to $L'$. We consider the potential $u(x)$ in $L$ to be a real, smooth and rapidly decaying function of $x$ and we look for
self-conjugate $\zeta$ and $\eta$ such that the new potential $u'(x)$ is also real and smooth. In addition, we require $\eta$ to be the left inverse
of $\zeta$, i.e., to obey the condition
\begin{equation}
 \eta\zeta =I,  \label{zeta2}
\end{equation}
so that
\begin{equation}
 L=\eta L'\zeta .  \label{LtL}
\end{equation}
In order to get a new potential $u'(x)$ not decaying along some directions of the plane, the two operator $L$ and $L'$ must be related by a
transformation more general than a similarity one. Therefore, we search for twisting operators having a product $\zeta\eta$, which is not equal to $I$, but which is given by
\begin{equation}
 \zeta\eta=I-P ,  \label{P}
\end{equation}
where $P$ is an orthogonal self-conjugate projector, since as a consequence of~(\ref{zeta2}) $P^{2}=P$ and, thanks to the self-conjugate property of $\zeta$
and $\eta$, $P^*=P$. The twisting operators $\zeta$ and $\eta$ generate a new potential $u'$ via~(\ref{Lzeta}), but also the new dressing operators
$\nu'$ and $\omega'$. In fact, taking into account~(\ref{Lnu}), we get by~(\ref{Lzeta}) $L'\zeta\nu =\zeta L\nu =\zeta\nu L_{0}$, and by~(\ref{Lnu})
$\omega\eta L'=\omega L\eta=L_{0}\omega\eta$. Therefore, the operators $\nu'$ and $\omega'$ defined by
\begin{equation}
 \nu'=\zeta\nu ,\qquad \omega'=\omega\eta, \label{tildenu1}
\end{equation}
obey equations
\begin{equation}
 L'\nu'=\nu'L_{0},\qquad \omega'L'=L_{0}\omega',  \label{tLnu}
\end{equation}
analogous to~(\ref{Lnu}) satisfied by the dressing operators $\nu$ and $\omega$.

Note that because of~(\ref{zeta2}) the scalar product of these dressing operators is equal to $I$ like for the original one in~(\ref{scalar}):
\begin{equation}
 \omega'\nu'=I.  \label{tildescalar}
\end{equation}
However, in contrast to $\nu$ and $\omega$, these operators do not satisfy a completeness relation, since, thanks to~(\ref{scalar}) and unlike it, one has
by~(\ref{P})
\begin{equation}
 \nu'\omega'+P=I.  \label{tcompl}
\end{equation}

In order to obtain a Darboux transformation we must specify the analyticity properties of the kernels $\zeta(p;\q)$ and $\eta(p;\q)$ of the twisting
operators with respect to the variables $\q$. Since, thanks to~(\ref{scalar}) and~(\ref{tildenu1}), we have $\zeta=\nu'\omega$ and $\eta =\nu\omega'$, the singularities of these kernels are related to the properties of $\nu'(p;\q)$ and $\omega'(p;\q)$. We require that
\begin{enumerate}
\item
 the kernels $\nu'(p;\q)$ and $\omega'(p;\q)$ are independent of $\q_{2}$ and have asymptotic behavior $\lim_{\q_{1}\to\infty}\nu'(p;\q_1)=\delta(p)$ and
$\lim_{\q_{1}\to\infty}\omega'(p;\q_1)=\delta(p)$,
\item the kernels $\nu'(p;\q_1)$ and $\omega'(p;\q_1)$ have
(correspondingly, right and left) simple poles with respect to the variable $\q_{1}$, i.e., there exist nontrivial limits
\begin{align}
 & \nu_{b_{l}}'(p)=\lim_{\q_{1}\rightarrow ib_{l}}(\q_{1}-ib_{l})\nu'(p;\q_1),  \label{nuresa} \\
 & \omega_{a_{j}}'(p)=\lim_{\q_{1}\rightarrow -p_{1}+ia_{j}}(\q_{1}+p_{1}-ia_{j})\omega'(p;\q_1),  \label{oresa}
\end{align}
where $a_{1},\ldots ,a_{N_{a}}$, $b_{1},\ldots ,b_{N_{b}}$ are $N_{a}+N_{b}$ parameters, which we choose to be all different, and real in order to
guarantee reality of the transformed potential,
\end{enumerate}
and, moreover, that
\begin{enumerate}\setcounter{enumi}{2}
\item the kernels $\zeta(p;\q)$ and $\eta (p;\q)$ besides the
discontinuities at $\q_{1\Im}=b_{l}$ and $\q_{1\Im}=a_{j}$, which follow from requirement 2, have no additional departures from analyticity with respect to $\q_{1}$.
\end{enumerate}

Then, the kernels of the operators $\zeta$, $\eta$ are given by the following equations
\begin{align}
 \zeta(p;\q_1)& =\delta(p)+\sum_{l=1}^{N_{b}}\int dp'\dfrac{\nu_{b_{l}}'(p-p')\omega(p';ib_{l}-p'_1)}{\q_{1}+p_1'-ib_{l}},  \label{expl1} \\
 \eta(p;\q_1)& =\delta(p)+\sum_{j=1}^{N_{a}}\int dp'\dfrac{\nu(p-p';ia_{j})\omega_{a_{j}}'(p')}{\q_{1}+p'_1-ia_{j}},  \label{expl2}
\end{align}
where notations~(\ref{nuresa}) and~(\ref{oresa}) were used. Let now $\chi'(x,\bk)$ and $\xi'(p;\bk)$ denote the functions defined in terms of the
dressing operators $\nu'$ and $\omega'$ in analogy with~(\ref{nuchi}) and let $\chi_{b_{l}}'(x)$ and $\xi_{a_{j}}'(x)$ be their residua
(cf.~(\ref{nuresa}) and~(\ref{oresa})). Then, it is easy to show that condition~(\ref{zeta2}) is equivalent to the set of equations
\begin{align}
 &\chi_{b_{l}}'(x)=-i\sum_{j=1}^{N_{a}}\chi (x,ia_{j})m_{jl}(x),\qquad l=1,\ldots ,N_{b},  \label{t-chi} \\
 &\xi_{a_{j}}'(x)=i\sum_{l=1}^{N_{b}}m_{jl}(x)\xi(x,ib_{l}),\qquad j=1,\ldots ,N_{a}.  \label{t-xi}
\end{align}
where $m(x)$ is the $N_{a}\times{N_{b}}$-matrix with elements
\begin{equation}
 m_{jl}(x)=\int\limits_{x_{1}}^{(a_{j}-b_{l})\infty}dy_{1}\,e^{(a_{j}-b_{l})(x_{1}-y_{1})}\xi_{a_{j}}'(y)\chi_{b_{l}}'(y)\Bigr|_{y_{2}=x_{2}},  \label{m3}
\end{equation}
which is well defined for bounded $\chi_{b_{l}}'$ and $\xi_{a_{j}}'$ and obeys
\begin{equation}
 \partial_{x_{1}}m_{jl}(x)=(a_{j}-b_{l})m_{jl}(x)-\xi_{a_{j}}'(x)\chi_{b_{l}}'(x).  \label{m5}
\end{equation}
In addition, from (\ref{tildeL}) and (\ref{Lzeta}), we get for the transformed potential $u'(x)$
\begin{equation}
 u'(x)=u(x)-2\partial_{x_{1}}\sum_{j=1}^{N_{a}}\sum_{l=1}^{N_{b}}\xi(x,ib_{l})\chi(x,ia_{j})m_{jl}(x).  \label{u'x}
\end{equation}

Then, due to~(\ref{expl1}) and~(\ref{expl2}), in order to define $\zeta$ and $\eta$ we must specify $\nu_{b_{l}}'$ and $\omega_{a_{j}}'$, i.e.,
$\chi_{b_{l}}'(x)$ and $\xi_{a_{j}}'(x)$. Thanks to~(\ref{t-chi}) and~(\ref{t-xi}), this means that we have to define the
matrix $m_{jl}(x)$. Inserting equations~(\ref{t-chi}) and~(\ref{t-xi}) into~(\ref{m5}) and using (\ref{dualJost}), we write the following
equation for the matrix $\widehat{m}_{jl}(x)=e_{}^{i(\ell (ia_{j})-\ell (ib_{l}))x}m_{jl}(x)$
\begin{equation}
 \partial_{x_{1}}\widehat{m}_{jl}(x)=-\sum_{j'=1}^{N_{a}}\sum_{l'=1}^{N_{b}}\widehat{m}_{jl'}(x)\Psi(x,ib_{l'})\Phi(x,ia_{j'})
 \widehat{m}_{j'l}(x). \label{m6}
\end{equation}
Omitting details, we present, here, directly its solution in the two equivalent forms
\begin{equation}
 \widehat{m}(x)=(E_{N_{a}}+c\Fo(x))^{-1}c=c(E_{N_{b}}+\Fo(x)c)^{-1},\label{m7}
\end{equation}
where  $c$ is an arbitrary real constant $N_{a}\times {N_{b}}$ matrix, $E_{N_{a}}$ and $E_{N_{b}}$ are the unity $N_{a}\times {N_{a}}$ and
$N_{b}\times{N_{b}}$ matrices, correspondingly, and $\Fo(x)$ is a $N_{b}\times {N_{a}}$ matrix with elements $\Fo_{lj}(x)=\Fo(x,ib_{l},ia_{j})$,
where
\begin{equation}
 \Fo(x,\bk,\bk')=\int\limits_{(\bk_{\Im}^{}-\bk_{\Im}')\infty}^{x_{1}}\!\!\!\!\!dx_{1}'\Psi (x',\bk)\Phi (x',\bk')\Bigl|_{x_{2}'=x_{2}^{}}.\label{Fkk}
\end{equation}

Now the potential~(\ref{u'x}) also can be written in the two forms
\begin{equation}
 u'(x)=u(x)-2\partial_{x_{1}}^{2}\ln \det (E_{N_{b}}+\Fo c)= u(x)-2\partial_{x_{1}}^{2}\ln \det (E_{N_{a}}+c\Fo).  \label{u'1}
\end{equation}
Inserting the matrix $m$ found above into (\ref{t-chi}), (\ref{t-xi}) and, then, the obtained equations into (\ref{expl1}), (\ref{expl2}), we derive
the explicit formulae for the dressing operators $\zeta$ and $\eta$. Finally, from~(\ref{tildenu1}) using equation~(\ref{dualJost}) and its analog
for the transformed (primed) Jost and dual Jost solutions we get for them
\begin{align}
 \Phi'(x,\bk)& =\Phi (x,\bk)-\Phi (x,ia)(E_{N_{a}}+c\Fo(x))^{-1}c\Fo(x,ib,\bk)=  \notag \\
 & =\Phi (x,\bk)-\Phi (x,ia)c(E_{N_{b}}+\Fo(x)c)^{-1}\mathcal{F}(x,ib,\bk),\label{PHI} \\
 \Psi'(x,\bk)& =\Psi (x,\bk)-\mathcal{F}(x,\bk,ia)(E_{N_{a}}+c\Fo(x))^{-1}c\Psi (x,ib)=  \notag \\
 & =\Psi (x,\bk)-\mathcal{F}(x,\bk,ia)c(E_{N_{b}}+\Fo(x)c)^{-1}\Psi (x,ib),\label{PSI}
\end{align}
where $j=1,\dots ,N_{a}$, $l=1,\dots ,N_{b}$, and
\begin{align*}
 & \Phi (x,ia)=\diag\{\Phi (x,ia_{j})\},\qquad \Psi(x,ib)=\diag\{\Psi(x,ib_{l})\}, \\
 & \Fo(x,\bk,ia)=\diag\{\Fo(x,\bk,ia_{j})\},\qquad \Fo(x,ib,\bk)=\diag\{\Fo(x,ib_{l},\bk)\},
\end{align*}
and where any of the two forms can be used. It is easy to see that both $\Phi'(x,\bk)$ and $\Psi'(x,\bk)$ have poles at $\bk=ib_l$ and
$\bk=ia_j$, correspondingly, and thanks to~(\ref{PHI}) and~(\ref{PSI}) we get that the residua of these functions are given in terms of their
values in the dual points by means of the relations
\begin{equation}
 \Phi'_{b_l}(x)=-i\sum_{j=1}^{N_a}\Phi'(x,ia_j)c_{jl},\qquad
 \Psi'_{a_j}(x)=i\sum_{l=1}^{N_b}c_{jl}\Psi'(ib_l),\label{PhiPsi:final}
\end{equation}
as expected. One can show that the potential $u'(x)$ in (\ref{u'1}) and the Jost solutions from (\ref{PHI}) and (\ref{PSI}) coincide with those
obtained in~\cite{BPPPr2001a}, including the case $N_{a}{\neq}N_{b}$ considered here, which is recovered by choosing some zero rows or columns in the constant matrix $C$ introduced in~\cite{BPPPr2001a}. We postpone to Sec.~\ref{soliton} the discussion on the regularity conditions for the potential given by~(\ref{u'1}).

\section{Resolvent}\label{resolvent}

Once the transformed operator $L'$ has been obtained, the operators $\zeta$ and $\eta $ can be used to investigate its spectral properties and
existence and uniqueness of the corresponding resolvent $M'$. Multiplying the first and the second equation in~(\ref{Lzeta}), respectively, from the
right by $\eta$ and from the left by $\zeta$, and recalling the definition~(\ref{P}) of $P$, we get the intertwining relation
\begin{equation}
 L'=\zeta L\eta +L_{\Delta},  \label{LMdelta}
\end{equation}
where we introduced
\begin{equation}\label{LDelta}
   L_{\Delta}=L'P=PL'.
\end{equation}
Thus we look for a resolvent in the form
\begin{equation}
M'=\zeta M\eta +M_{\Delta},\label{Mtildedelta}
\end{equation}
where $\zeta{M}\eta$ is determined by the above construction. Indeed, due to~(\ref{LM}) and~(\ref{scalar}) we deduce that
\begin{equation}
 \zeta{M}\eta =\nu'M_{0}\omega',  \label{AP380}
\end{equation}
and, then, we get for its kernel in the $x$-space the following bilinear expression in terms of the transformed Jost solutions
\begin{align}
 (\zeta M\eta)(x,x';q)& =-\sgn(x_{2}-x_{2}')\dfrac{e^{-q(x-x')}}{2\pi}\int \!\!dp_{1}
 \theta \left((q_{2}+p_{1}^{2}-q_{1}^{2})(x_{2}-x_{2}')\right) \times  \notag \\
 & \times \Phi'(x;p_{1}+iq_{1})\Psi'(x';p_{1}+iq_{1}).\label{zetaMeta}
\end{align}

The second term in~(\ref{Mtildedelta}), $M_{\Delta}$, is to be determined. By using~(\ref{Lzeta}), (\ref{M}), and~(\ref{P}), we deduce that $M'$
in~(\ref{Mtildedelta}) is the right or left inverse of $L'$ iff $M_{\Delta}$ is a solution, respectively, of the first or the second of the operator equations
\begin{equation}
L'M_{\Delta}=P, \qquad M_{\Delta}L'=P.\label{iff}
\end{equation}
Below we consider the solvability of these equations in the case of pure soliton potentials. In order to complete the discussion of the generic case, we mention that an explicit form of the operator $P$ defined in~(\ref{P}) can be derived by inserting there $\zeta$ and $\eta$ given
in~(\ref{expl1}) and~(\ref{expl2}). Thanks again to~(\ref{dualJost}), we get for the hat-kernel of this operator the expression
\begin{equation}
 \widehat{P}(x,x';q)=i\delta (x_{2}-x_{2}')\sum_{n=1}^{\Ncal}\theta (q_{1}-\alpha_{n})\res_{\bk=i\alpha_{n}}\Phi'(x,\bk)\Psi'(x',\bk).  \label{AP4}
\end{equation}
Here we introduced the set of parameters
\begin{equation}
 \{\alpha_{1},\ldots ,\alpha_{\Ncal}\}=\{a_{1},\ldots ,a_{N_{a}},b_{1}\ldots ,b_{N_{b}}\},\qquad \Ncal=N_{a}+N_{b},\label{am}
\end{equation}
in order to make explicit the symmetry properties with respect to the parameters $a_{j}$, $b_{l}$ used above. This kernel is different from zero only
in the interval where $q_1$ is between the lowest and highest values of the $\alpha_m$'s. Indeed, while for $q_1$ below this interval this is obvious
by~(\ref{AP4}), for the values of $q_1$ above the interval this follows from the equality
\begin{equation}
 \sum_{n=1}^{\Ncal}\res_{\bk=i\alpha_{n}}\Phi'(x,\bk)\Psi'(x',\bk)=0,\label{zero}
\end{equation}
that in its turn is a consequence of~(\ref{PhiPsi:final}). Now, according to the discussion above, we expect that
$\widehat{M}_{\Delta}$ has a structure similar to $\widehat{P}(x,x';q)$. It is clear that the kernel
\begin{equation}
 \widehat{M}_{\Delta}(x,x';q)=\mp i\theta\bigl(\pm(x_{2}-x_{2}')\bigr)
 \sum_{n=1}^{\Ncal}\theta (q_{1}-\alpha_{n})\res_{\bk=i\alpha_{n}}\Phi'(x,\bk)\Psi'(x',\bk),\label{hatMd}
\end{equation}
obeys equations $\Lo'_{x}\widehat{M}_{\Delta}(x,x';q)=P(x,x';q)$, $\Lo'^{\text{d}}_{x'}\widehat{M}_{\Delta}(x,x';q)=P(x,x';q)$ for any sign. But it
is easy to see that, in general, the kernel $M_{\Delta}(x,x';q)$ constructed by means of~(\ref{7}) is growing at space infinity and cannot be the kernel of an extended operator as defined in Sec.~\ref{back}. Below we investigate this problem in detail by means of a specific example.

\section{Pure soliton potential and Jost solutions}\label{soliton}

In the case $u(x)\equiv 0$ one gets the general $N$-soliton solution, where $N=\max \{N_{a},N_{b}\}$. The transformed potential $u'(x)$ and the
corresponding Jost solutions can be easily obtained from the general expression derived in Sec.~(\ref{twist}). Thus, the transformed potential is given by any of the formulae in~(\ref{u'1}) where now the matrix $\Fo(x)$ has elements
\begin{equation}
 \Fo_{lj}(x)=\dfrac{e^{(a_j-b_l)(x_1+(a_j+b_l)x_2)}}{a_j-b_l}.\label{Flj}
\end{equation}
By expanding the determinants on the r.h.s.\ of (\ref{u'1}) and by using the Binet--Cauchy formula one gets that, in order to have a regular potential, it is
sufficient that the real matrix $c$ satisfies the following characterization requirements (equivalent to those in~\cite{BC})
\begin{equation}
 \Lambda\left(\begin{array}{c}
 l_{1},l_{2},\dots ,l_{n} \\
 j_{1},j_{2},\dots ,j_{n}\end{array}\right)
 c\left(\begin{array}{c}
 j_{1},j_{2},\dots ,j_{n} \\
 l_{1},l_{2},\dots ,l_{n}
 \end{array}\right) \geq 0,\qquad \Lambda_{lj}=\dfrac{1}{a_j-b_l}  \label{condition}
\end{equation}
for any $1\leq n\leq \min \{N_{a},N_{b}\}$ and all minors, i.e., any choice of $1\leq l_{1}<l_{2}<\dots <l_{n}\leq N_{b}$ and $1\leq j_{1}<j_{2}<\dots
j_{n}\leq N_{a}$. Here, for the minors we used the standard notation
\begin{equation}
A\left(\begin{array}{c}
j_{1},j_{2},\dots ,j_{n} \\
l_{1},l_{2},\dots ,l_{n}
\end{array}\right) =\det \left\vert \left\vert
\begin{array}{cccc}
a_{j_{1}l_{1}} & a_{j_{1}l_{2}} & \cdots & a_{j_{1}l_{n}} \\
a_{j_{2}l_{1}} & a_{j_{2}l_{2}} & \cdots & a_{j_{2}l_{n}} \\
\cdots & \cdots & \cdots & \cdots \\
a_{j_{n}l_{1}} & a_{j_{n}l_{2}} & \cdots & a_{j_{n}l_{n}}
\end{array}\right\vert \right\vert .
\end{equation}
Formulae~(\ref{PHI}) and~(\ref{PSI}), after rather cumbersome calculations that we skip here, can be more simply expressed as ratios of
determinants.  As in (\ref{u'1}) one can alternatively, but equivalently, use determinants of $(N_{b}\times{N_{b}})$- or
$(N_{a}\times{N_{a}})$-matrices. We use below the last case and we use the symmetric notation~(\ref{am}). In these terms for the functions $\chi'$
and $\xi'$ related the Jost solutions by~(\ref{dualJost}) we get
\begin{align}
 \chi'(x,\bk)& =\left(\,\,\prod_{l=1}^{N_{b}}(b_{l}+i\bk)^{-1}\right)\dfrac{\tau_{\chi}(x,\bk)}{\tau (x)},  \label{sol19:1} \\
 \xi'(x',\bk)& =\left(\,\,\prod_{l=1}^{N_{b}}(b_{l}+i\bk)\right) \dfrac{\tau_{\xi}(x',\bk)}{\tau (x')}, \label{sol19:2}
\end{align}
where the $\tau$ functions are determinants defined by
\begin{align}
 & \tau_{\chi}(x,\bk)=\det \bigl(\mathfrak{A}\,e^{\Ao(x)}(\alpha +i\bk)\Do\bigr) ,  \label{tauchi} \\
 & \tau_{\xi}(x',\bk)=\det \bigl(\mathfrak{A}\,e^{\Ao(x')}(\alpha +i\bk)^{-1}\Do\bigr) ,  \label{tauxi} \\
 & \tau (x)=\det \bigl( \mathfrak{A\,}e^{\Ao(x)}\Do\bigr) ,\label{tau}
\end{align}
i.e., determinants of $(N_{b}\times N_{b})$-matrices obtained as products of square and rectangular matrices. Precisely,
\begin{align}
 & \mathfrak{A}_{ln}=\alpha_{n}^{N_{b}-l},\qquad \alpha +i\bk=\diag\{\alpha_{n}+i\bk\} \\
 & e^{\Ao(x)}=\diag\{e^{\Ao_{n}(x)}\},\qquad \Ao_{n}(x)=\alpha_{n}x_{1}+\alpha_{n}^{2}x_{2} \\
 & \Do=\left(\begin{array}{l}
 d \\E_{N_{b}}\end{array}\right) ,\qquad d_{jl}=c_{jl}\dfrac{\prod_{l'=1,\,l'\neq l}^{N_{b}}(b_{l}-b_{l'})}{\prod_{l'=1}^{N_{b}}(a_{j}-b_{l'})},\label{D}
\end{align}
where $l=1,2,\ldots ,N_{b}$, $j=1,2,\dots,N_{a}$, $n=1,2,\ldots ,\Ncal$. As far as the potential is concerned, it can be expressed as
$u'(x)=-2\partial_{x_{1}}^{2}\log\tau (x)$ and, taking into account~(\ref{tau}), one recovers the expression obtained in~\cite{BK,BC,ChK1} using the
tau functions.

\section{$N$-soliton solutions in the case $N_b=1$}\label{1ray}

Now, we restrict ourselves to the special case of the $N$-soliton potential with $N_{b}=1$ and $N_{a}=N$ arbitrary. In
addition, for simplicity, without loss of generality, we choose $\alpha_{1}<\alpha_{2}<\dots <\alpha_{\Ncal}$. In this situation the potential $u'(x)$
has on the $x$-plane $N_a$ ingoing rays and one outgoing ray, as schematically shown in Fig.~1. In this case
$\tau$-functions~(\ref{tauchi})--(\ref{tau}) take the simple form
\begin{align}\label{tau1}
  \tau_{\chi}(x,\bk)&=\sum_{m=1}^{\Ncal}f_me^{\Ao(x)}(\alpha_m+i\bk),\qquad\tau_{\xi}(x,\bk)=
  \sum_{m=1}^{\Ncal}\dfrac{f_me^{\Ao(x)}}{\alpha_m+i\bk},\nonumber\\
  \tau(x)&=\sum_{m=1}^{\Ncal}f_me^{\Ao(x)},
\end{align}
where the $f_m$'s are obtained from the elements $d_{m1}$'s of matrix $d$ in~(\ref{D}) by a permutation which take into account the chosen ordering of the $\alpha_m$'s. Condition~(\ref{condition}) here means that all $f_m>0$.
\begin{figure}[th]
\begin{center}
\includegraphics[width=5cm,height=3cm]{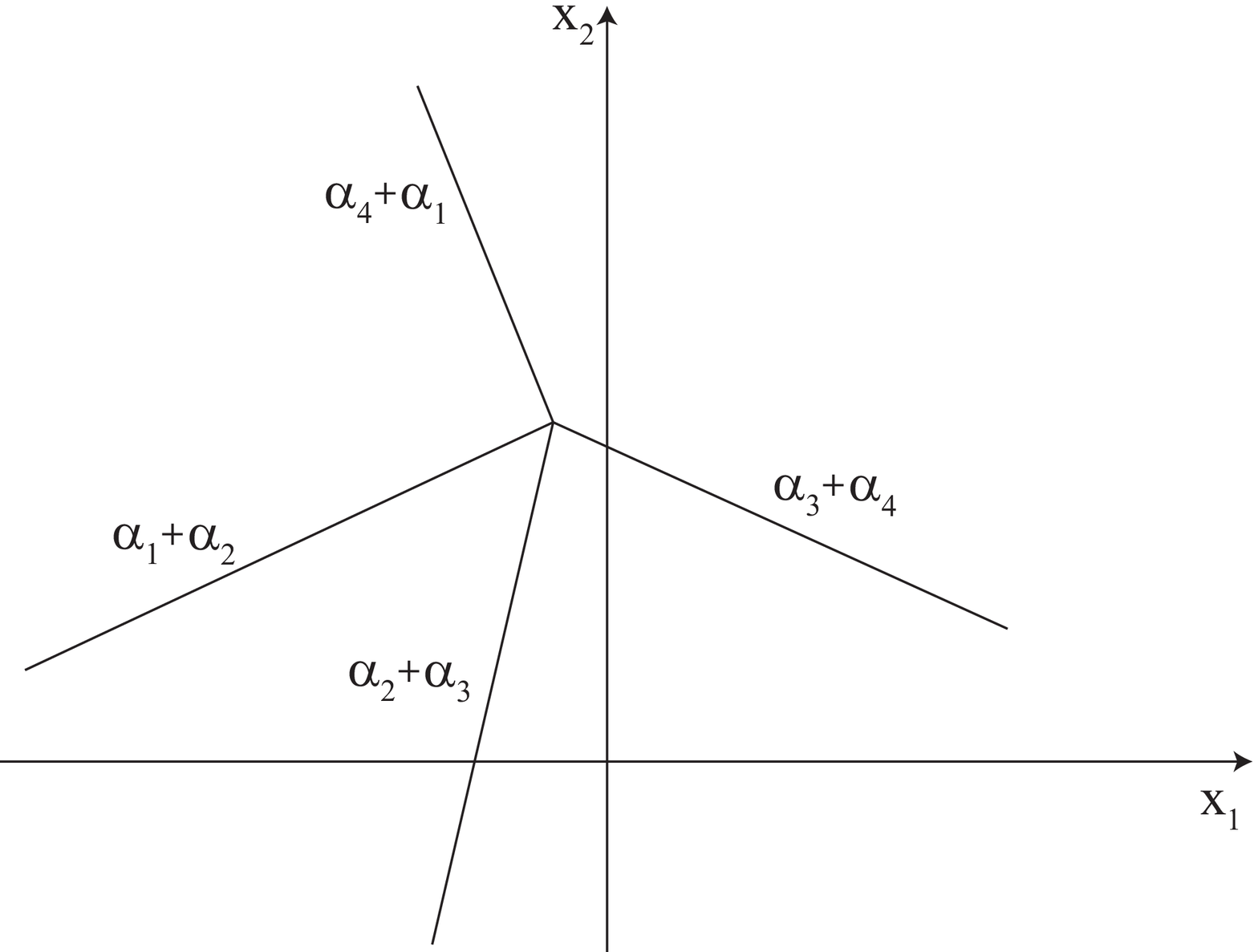}
\includegraphics[width=5cm,height=3cm]{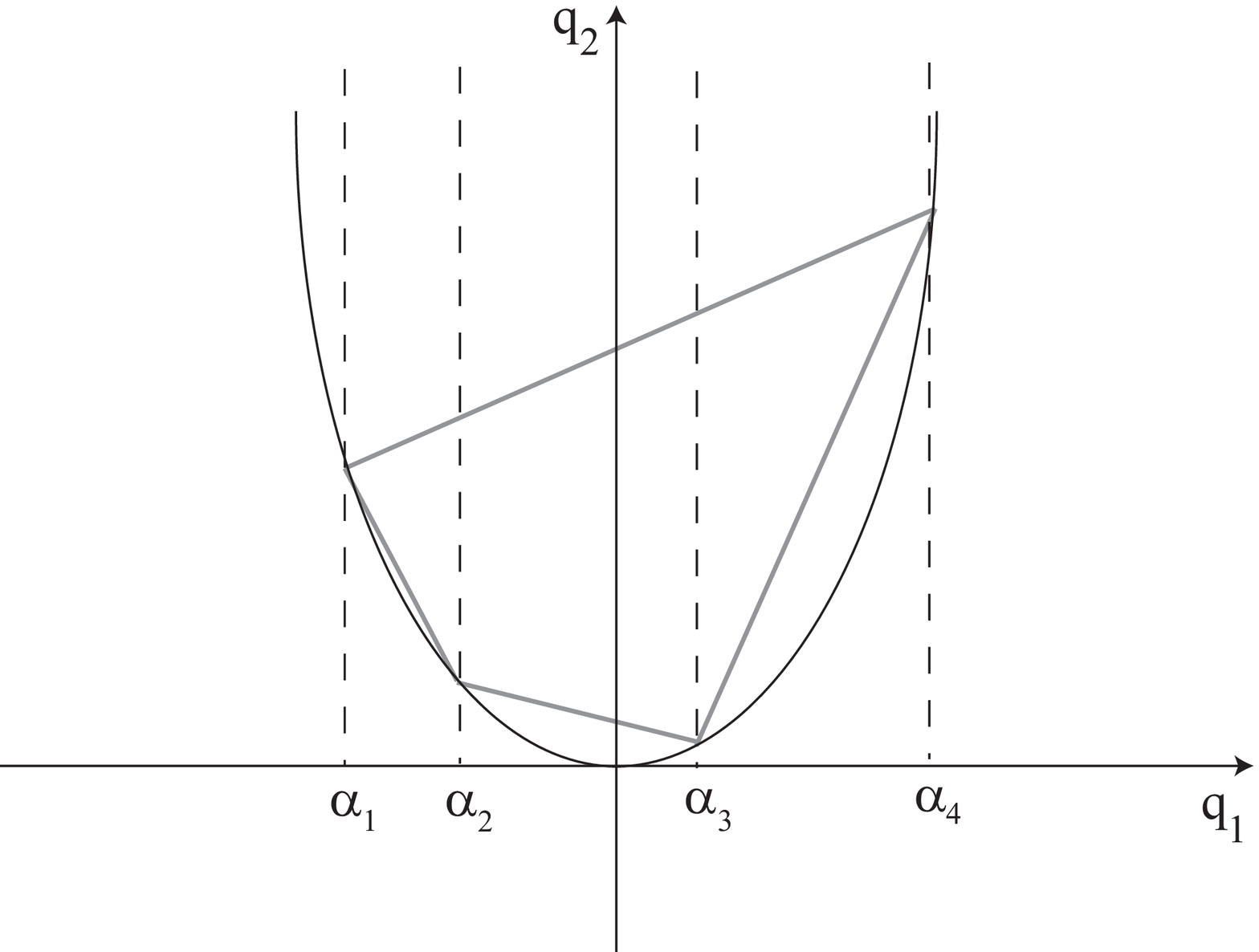}
\end{center}
\caption{Rays and Polygon for $\Ncal=4$}
\end{figure}

We want to show that the extended operator $L'(q)$ corresponding to this potential, as announced in the introduction, has a left self-conjugate
annihilator $K(q)$ for $q$ belonging to a domain of the $q$-plane. Precisely, let us consider the domain inside the polygon inscribed in the
parabola $q_2^{}=q_1^2$ of the $q$-plane and with vertices at points $q_1=\alpha_m$, $m=1,\ldots,N$, whose
characteristic function is given by
\begin{align}
 \kappa(q)& =\sum_{m=1}^{\Ncal-1}[\theta (q_{1}-\alpha_{m+1})-\theta(q_{1}-\alpha_{m})]\times\notag \\
 & \times \lbrack \theta (q_{2}-(\alpha_{m}+\alpha_{n})q_{1}+\alpha_{m}\alpha_{n})-
 \theta (q_{2}-(\alpha_{1}+\alpha_{\Ncal})q_{1}+\alpha_{1}\alpha_{\Ncal}).\label{cq}
\end{align}
Notice that this polygon can be considered as dual to the ray structure (see Fig.~1 for the case $\Ncal=4$).

The main remark needed in order to get the annihilator $K(q)$ is that the function
\begin{equation}
\psi(x;q)=\dfrac{\kappa(q)e^{qx}}{\tau(x)}\label{psi}
\end{equation}
is bounded in the $x$-plane. More precisely, as one can prove after a detailed study, it is exponentially decaying for $x$ going to infinity in any
direction on the plane, for $q$ inside the polygon, and, for $q$ on the borders of the polygon, it has directions of nondecaying (but bounded)
behavior.

Then, since thanks to~(\ref{dualJost}), (\ref{sol19:1}) and~(\ref{tau1}) $1/\tau$ is proportional to the value of the dual Jost solution
$\Psi'(x,\bk)$ at $\bk=ib_1$, we have
\begin{equation}\label{L'psi}
    {\Lo'_{x'}}^{\text{d}}\widehat\psi(x';q)=0,
\end{equation}
and, consequently,
\begin{equation}\label{KL'}
    KL'=0
\end{equation}
for any operator $K$ with kernel
\begin{equation}\label{K}
    K(x,x';q)=\varphi(x;q)\psi(x';q)
\end{equation}
where $\varphi(x;q)$ is any arbitrary self-conjugate function, bounded in $x$ and identically zero outside the polygon introduced above.

One can verify directly that $KP=K$. If, in addition, we choose $\varphi(x;q)$ of the form $\varphi(x;q)=(P\gamma)(x;q)$ with $\gamma(x;q)$ a
self-conjugate function, bounded in $x$ and identically zero outside the polygon, and such that
\begin{equation}\label{gammapsi}
    \int dx\,\gamma(x;q)\psi(x,q)=\kappa(q),
\end{equation}
one also get
\begin{equation}\label{phipsi}
    P\varphi=\varphi,\qquad\int dx\,\varphi(x;q)\psi(x,q)=\kappa(q).
\end{equation}
Then, it is easy to check that $K$ is a self-conjugate projector commuting with the projector $P$, i.e., we have
\begin{equation}\label{KP}
    K^*=K,\qquad K^2=K,\qquad PK=KP=K.
\end{equation}
The existence of this annihilator proves that the operator $L'$ cannot have right inverse for $q$ belonging to the polygon defined by the
characteristic function~(\ref{cq}). In order to prove that, on the contrary, the resolvent $M'(q)$ exists for $q$ outside the polygon we go back to the hat-kernel $\widehat{M}_{\Delta}(x,x';q)$ defined in~(\ref{hatMd}) and we study the boundedness properties of  $M_{\Delta}(x,x';q)=e^{-q(x-x')}\widehat{M}_{\Delta}(x,x';q)$. Since, thanks to~(\ref{zero}), this function is identically equal to zero
outside the strip $\alpha_1<q_1<\alpha_{N+1}$ on the $q$-plane, we need to investigate its properties only in this strip. Using~(\ref{dualJost}), (\ref{sol19:1}), (\ref{sol19:2}), and~(\ref{tau1}) we rewrite it
explicitly as
\begin{align}
 M_{\Delta}(x,x';q)& =\dfrac{\pm e^{-q(x-x')}\theta\bigl(\pm(x_2-x'_2)\bigr)}{\tau (x)\tau (x')}\times\nonumber\\
 &\times\sum_{m,n=1}^{\Ncal}f_mf_n\theta(q_{1}-\alpha_{m})
 (\alpha_{m}-\alpha_{n})e^{\Ao_{m}(x)+\Ao_{n}(x)}. \label{Md}
\end{align}
Then, one can prove that, choosing the upper (bottom) sign for $q$ above (below) the polygon in the
strip $\alpha_1<q_1<\alpha_{N+1}$, the kernel $M_{\Delta}(x,x';q)$ is a bounded function of $x$ and $q$ and it defines the kernel of an extended operator $M_{\Delta}(q)$ according to the definition in Sec.~\ref{back}. We conclude that, for $q$ outside the polygon, $M_{\Delta}(q)$ satisfy equations (\ref{iff}) and, consequently, the resolvent $M'(q)$ exists and is given by~(\ref{Mtildedelta}).

Notice that the
total Green's function $G(x,x',\bk)$ of the operator $\Lo'$ can be defined (see~\cite{BPPPr2001a,BPPPr2002}) as the value of the kernel $M'(x,x';q)$ at
$q_1=\bk_{\Im}$, $q_2=\bk_{\Im}^2-\bk_{\Re}^2$ for a complex spectral parameter $\bk=\bk_{\Re}+i\bk_{\Im}$. These values of $q$ lie outside the
parabola $q_{2}=q_{1}^{2}$, thus outside the polygon and touch it at the vertices only. Then the Green's function exists for any $\bk$ but it is
singular at the points $\bk=i\alpha_m$ corresponding to the vertices of the polygon. These are the only singularities of the Green's function, since the
discontinuities of the first term in~(\ref{Mtildedelta}) at $q_1=\alpha_m$ (see~(\ref{zetaMeta})) are compensated (outside the polygon) by the
discontinuities of the second term, as follows from~(\ref{hatMd}).

Finally, it is worth noting that the case $N_a=1$ and $N_b$ arbitrary can be handled in a analogous way, having the operator $L'(q)$ a right instead of a left annihilator for $q$ inside the polygon, and that in the case $N_a=N_b=1$ the polygon reduces to the segment of the line $q_2=(\alpha_1+\alpha_2)q_1-\alpha_1\alpha_2$ with end points $q_1=\alpha_1$ and $q_1=\alpha_2$. In this case one recovers the results obtained in~\cite{BPPPr2001a}.

\bigskip
\textbf{Acknowledgments}
This work is supported in part by the grant RFBR--CE \# 06-01-92057, grant NWO--RFBR  \# 047.011.2004.059, grant RFBR \# 08-01-00501, Scientific Schools 795.2008.1, by the Program of RAS ''Mathematical Methods of the Nonlinear Dynamics'', by INFN and by Consortium E.I.N.S.T.E.IN. AKP thanks Department of Physics of the University of Salento (Lecce) for kind hospitality.


\begin{thebibliography}{99}
\frenchspacing
\bibitem{D1974} V.~S. Dryuma, \textsl{Sov. JETP Lett.} \textbf{19} 387--388 (1974).
\bibitem{ZS1974} V. E. Zakharov and A. B. Shabat, \textsl{Func. Anal. Appl.} \textbf{8} 226--235 (1974).
\bibitem{AYF1983} M. J. Ablowitz, D. Bar Yacoov and A. S. Fokas, \textsl{Stud. Appl. Math.} \textbf{69} 135--143 (1983).
\bibitem{Grinev1988} G. Grinevich and P. S. Novikov, \textsl{Funct. Anal. Appl} \textbf{22} 19--27 (1988).
\bibitem{BPPPo1992a} M.~Boiti, F.~Pempinelli, A.~K.~Pogrebkov and M.~C.~Polivanov, \textsl{Inverse Problems} \textbf{8} 331--364 (1992).
\bibitem{BPPPo1992c} M.~Boiti, F.~Pempinelli, A.~K.~Pogrebkov and M.~C.~Polivanov, \textsl{Theor. Math. Phys.} \textbf{93} 1200--1224 (1992).
\bibitem{BPP1994c} M.~Boiti, F.~Pempinelli and A.~K. Pogrebkov, \textsl{J. Math. Phys.} \textbf{35} 4683--4718 (1994).
\bibitem{BPP1997} M.~Boiti, F.~Pempinelli and A. K. Pogrebkov, \textsl{Inverse Problems} \textbf{13} L7--L10 (1997).
\bibitem{BPPPr1998} M.~Boiti, F.~Pempinelli, A.~K.~Pogrebkov and B.~Prinari, \textsl{Theor. Math. Phys.} \textbf{116} 741--781 (1998).
\bibitem{Pr2000} B. Prinari, \textsl{Inverse Problems} \textbf{16} 589--603 (2000).
\bibitem{BPPPr2001a} M.~Boiti, F.~Pempinelli, A.~Pogrebkov and B. Prinari, \textsl{Inverse Problems} \textbf{17} 937--957 (2001).
\bibitem{BPPPr2002} M.~Boiti, F.~Pempinelli, A.~K.~Pogrebkov and B.~Prinari, \textsl{J. Math. Phys.} \textbf{43} 1044--1062 (2002).
\bibitem{BPPPr2005a} M.~Boiti, F.~Pempinelli, A.~K.~Pogrebkov and B.~Prinari, \textsl{Theor. Math. Phys.} \textbf{144}, 1100--1116 (2005).
\bibitem{BPPPr2005b} M.~Boiti, F.~Pempinelli, A.~K.~Pogrebkov and B.~Prinari, \textsl{Proc. Steklov Inst. Math.} \textbf{251} 6--48 (2005).
\bibitem{Villaroel} J.~Villarroel and M.J.~Ablowitz, \textsl{Stud. Appl. Math.} \textbf{109} 151--162 (2002).
\bibitem{BPP2006a} M.~Boiti, F.~Pempinelli and A.~K.~Pogrebkov, \textsl{J. Phys. A: Math Gen.} \textbf{39} 1877--1898 (2006).
\bibitem{BPP2006b} M.~Boiti, F.~Pempinelli and A.~K.~Pogrebkov, \textsl{J. Math. Phys.} \textbf{47} 123510 1--43 (2006).
\bibitem{MZBIM1977} S.~V.~Manakov, V.~E.~Zakharov, L.~A.~Bordag, A.~R.~Its, and V.~B.~Matveev, \textsl{Phys. Lett.} \textbf{A 63} 205--206 (1977).
\bibitem{BK} G. ~Biondini and Y. ~Kodama, \textsl{J. Phys. A: Math. Gen.} \textbf{36} 10519--10536 (2003).
\bibitem{BC} G.~Biondini and S.~Chakravarty, \textsl{J. Math. Phys.} \textbf{47} 033514 1--26 (2006).
\bibitem{BC2} G.~Biondini, \textsl{Phys. Rev. Lett.} \textbf{99} 064103 1--4 (2007).
\bibitem{ChK1} S.~Chakravarty and Y.~Kodama, \textsl{J. Phys. A: Math. Theor.} \textbf{41} 275209 (2008).
\end{thebibliography}
\end{document}